\documentclass[fleqn,usenatbib]{mnras}

\usepackage{newtxtext,newtxmath}

\usepackage[T1]{fontenc}

\DeclareRobustCommand{\VAN}[3]{#2}
\let\VANthebibliography\thebibliography
\def\thebibliography{\DeclareRobustCommand{\VAN}[3]{##3}\VANthebibliography}

\usepackage{graphicx}	
\usepackage{amsmath}	
\usepackage{xcolor}
\usepackage{subfig}

\def\sw{Swift~J004929.5-733107}

\newcommand{\swift}{\textit{Swift}}
\newcommand{\bexrb}{BeXRB}
\newcommand{\bexrbs}{BeXRBs}

\title[\sw]{The Be/neutron star system \sw{} in the Small Magellanic Cloud - X-ray characteristics and optical counterpart candidates}

\author[M. J. Coe et al.]{M.~J. Coe$^{1}$\thanks{E-mail: mjcoe@soton.ac.uk},
J.~A. Kennea$^{2}$, P.~A. Evans$^{3}$,
L.~J. Townsend$^{4,5}$, A. Udalski$^{6}$, 
I.~M. Monageng$^{4,7}$ and 
\newauthor
D.~A.~H. Buckley$^{4,5}$
\\
$^{1}$Physics \& Astronomy, The University of Southampton, SO17 1BJ, UK\\
$^{2}$Department of Astronomy and Astrophysics, The Pennsylvania State University, 525 Davey Lab, University Park, PA 16802, USA\\
$^{3}$University of Leicester, Astronomy Research Group, School of Physics \& Astronomy, University Road, Leicester LE1 7RH, UK\\
$^{4}$South African Astronomical Observatory, PO Box 9, Observatory, 7935, Cape Town, South Africa\\
$^{5}$Southern African Large Telescope, PO Box 9, Observatory, 7935, Cape Town, South Africa\\
$^{6}$Astronomical Observatory, University of Warsaw, Al. Ujazdowskie 4, 00-478 Warszawa, Poland\\
$^7$Department of Astronomy, University of Cape Town, Private Bag X3, Rondebosch 7701, South Africa\\
}

\date{}

\pubyear{2021}

\begin{document}
\label{firstpage}
\pagerange{\pageref{firstpage}--\pageref{lastpage}}
\maketitle

\begin{abstract}

\noindent \sw{} is an X-ray source in the Small Magellanic Cloud (SMC) that has been reported several times, but the optical counterpart has been unclear due to source confusion in a crowded region of the SMC. Previous works proposed [MA93] 302 as the counterpart, however we show here, using data obtained from the S-CUBED project, that the X-ray position is inconsistent with that object. Instead we propose a previously unclassified object which has all the indications of being a newly identified Be star exhibiting strong H$\alpha$ emission. 
Evidence for the presence of significant I-band variability strongly suggest that this is, in fact, a Be type star with a large circumstellar disk. Over 18 years worth of optical monitoring by the OGLE project reveal a periodic modulation at a period of 413d, probably the binary period of the system. A SALT optical spectrum shows strong Balmer emission and supports a proposed spectral classification of B1-3 III-IVe. The X-ray data obtained from the S-CUBED project reveal a time-averaged spectrum well fitted by a photon index $\Gamma = 0.93\pm 0.16$. Assuming the known distance to the SMC the flux corresponds to a luminosity $\sim10^{35}$ erg-s$^{-1}$. All of these observational facts suggest that this is confirmed as a Be star-neutron star X-ray binary (\bexrb) in the SMC, albeit one with an unusually long binary period at the limits of the Corbet Diagram.

\end{abstract}




\section{Introduction}

\bexrb\ are a large sub-group of the well-established category of High Mass X-ray Binaries (HMXB) characterised by being a binary system consisting of a massive mass donor star, normally an OBe type, and an accreting compact object, a neutron star. They are particularly prevalent in the Small Magellanic Cloud (SMC) which contains the largest known collection of \bexrbs\ . ~Whilst catalogues have been produced listing such systems in the SMC (for example \cite{ck2015, hs2016}), it clear that there are many more systems yet to be identified. This lack of completeness primarily arises from the transient nature of their X-ray emission, but also sometimes from an inaccurate X-ray position making the identity of the optical counterpart difficult in crowded stellar fields like the SMC. Since the identity and numbers of HMXBs are important tools in our understanding of star formation in a low metalicity environments like the SMC \citep{sg2005}, it is important to continue the search for such systems.

In this paper we report on both X-ray and optical recent photometry of the X-ray system \sw{}. This X-ray source has been previously reported several times, see for example \cite{sasaki2000,hep2008, hs2016,kennea2018}. 
The possible association with the nearby proposed Be star [MA93] 302 \citep{ma93} was first suggested by \cite{hs2000} and reiterated in subsequent publications.

In this work new observations are reported of this X-ray object as part of the S-CUBED regular monitoring of the SMC \citep{kennea2018} using the {\it Neil Gehrels Swift Observatory} \citep{gehrels04}. By combining all the many detections it has been possible to produce an accurate source location and X-ray lightcurve. The new position excludes the previous proposed optical counterpart [MA93] 302 and instead points to another nearby star. This star has not been previously identified nor classified, partially because of some source confusion with a third nearby star. To resolve this confusion optical spectra of all three stars were obtained under excellent seeing conditions with the Southern African Large Telescope (SALT). These spectra combined with the new accurate X-ray position resolve the previous ambiguity over which star is the correct counterpart to \sw{}.

In addition historical optical photometric data are reported here from the OGLE project \citep{Udalski2015} showing evidence for a probable binary period of 413d. Furthermore, significant (V--I) colour changes are reported which are suggested to be related to the changing circumstances in a circumstellar disk.
Therefore it is proposed here that this is the correct optical counterpart to \sw{} and it is a newly identified Be star in the SMC.

\section{Observations}

\subsection{S-CUBED and \swift\ Archival Observations}

\sw{} was detected by the S-CUBED survey \citep{kennea2018}, a shallow weekly X-ray survey of the optical extent of the SMC by the Swift X-ray Telescope (XRT; \citealt{burrows05}). Individual exposures in the S-CUBED survey are typically 60s long, and occur weekly, although interruptions can occur due to scheduling constraints. Starting with the S-CUBED observation taken on MJD 57722 (29 Nov 2016), S-CUBED detected it on many subsequent occasions and internally numbered it SC404.

\begin{figure}

	\includegraphics[width=8cm,angle=-00]{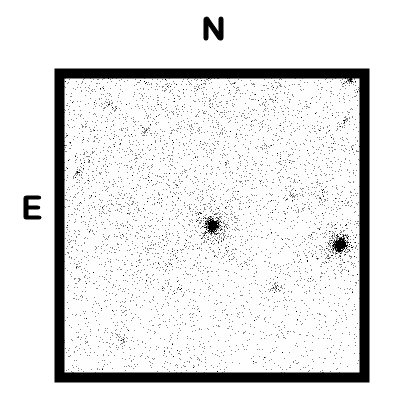}
    \caption{A deep S-CUBED stacked X-ray image of the field of \sw{} which is centred at a position of RA(2000)=00h 49m 30.49s, Dec(2000)=-73$^{\circ}$ 31' 09.3". The central object is the subject of this paper and is consistent with a single point source. The size of the field shown is 12 x 12 arcmin. The source near the western edge of the field is the symbiotic star system RX J0048.4-7332 \citep{orio2007}.}
    \label{fig:fcx}
\end{figure}

Through its lifetime, \emph{Swift} has observed the location of this source on many occasions in addition to the S-CUBED survey, giving a total of 96 ks of Photon Counting (PC) mode exposure at the source location. Using the online XRT analysis tools\footnote{\url{https://www.swift.ac.uk/user\_objects}} \citep{Evans09} a
spectrum and a position were obtained by stacking all of these data. The resulting stacked X-ray image is shown in Fig \ref{fig:fcx}. The best position was the `astrometric position', which was obtained by aligning XRT field sources with 2MASS objects (see \citealt{Evans14} for details), and is:\\

RA(2000)=00h 49m 30.49s, Dec(2000)=-73$^{\circ}$ 31' 09.3", error radius 1.3" (90\% confidence limit)\\

We note that this position was covered by a \emph{Chandra} observation on 2014 April 6 (PI: Predehl). Examination of these data show that SC404 was detected, at a position that is consistent with this reported S-CUBED position to within 0.5".

The summed PC mode X-ray spectrum is well described by an absorbed power-law model. With $N_H$ fixed at a standard value of $5.2\times10^{21}\ \mathrm{cm}^{-2}$ \cite{Willingale2013}, a photon index of $\Gamma = 0.93\pm 0.16$ with a reduced $\chi^2 = 0.89$ (14 degrees of freedom) is found. It is noted that this hard X-ray spectrum is consistent with other \bexrbs\ in the SMC, and given that the companion is a Be-star (see below), this is highly suggestive that \sw{} is, in fact, a \bexrb\ system.

Assuming a standard SMC distance of 62~kpc \citep{scowcroft2016} and correcting for absorption fixed at the value derived from \cite{Willingale2013}, this corresponds to a 0.5-10~keV luminosity of $(9.8\pm 1.7) \times 10^{34}$~erg-$s^{-1}$. 

S-CUBED observations taken with the UV/Optical Telescope (UVOT; \citealt{Roming05}) were also analysed, utilizing the standard \texttt{uvotmaghist} tool. In addition to S-CUBED observations we also analyzed UVOT data taken in archival observations of the field in the same way. Note that S-CUBED UVOT data is taken entirely utilizing the \textit{uvw1} filter, and in order to compare the the brightness of the source in archival data with those in S-CUBED, we only present observations taken utilizing that filter. 

\begin{figure}

	\includegraphics[width=8cm,angle=-00]{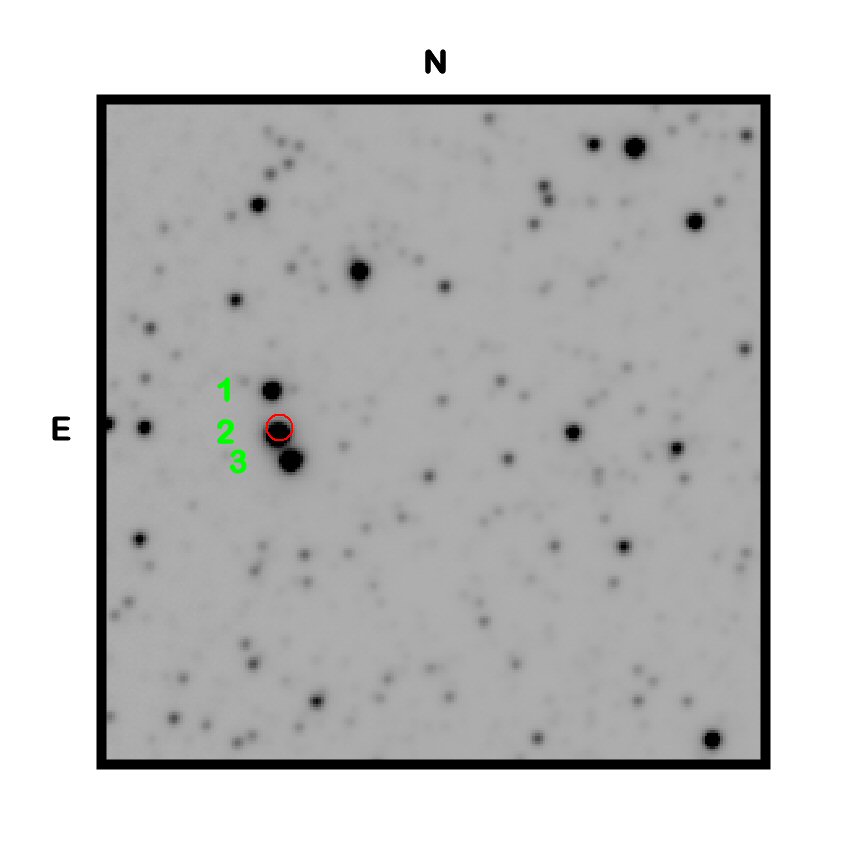}
    \caption{OGLE I band image of a 66.5 x 66.5 arcsecond field with the three nearby optical objects to \sw{} identified. The 90\% uncertainty of the X-ray position of \sw{} is shown by the circle.}
    \label{fig:fc}
\end{figure}

\subsection{OGLE}

\begin{figure*}

	\includegraphics[width=16cm,angle=-0]{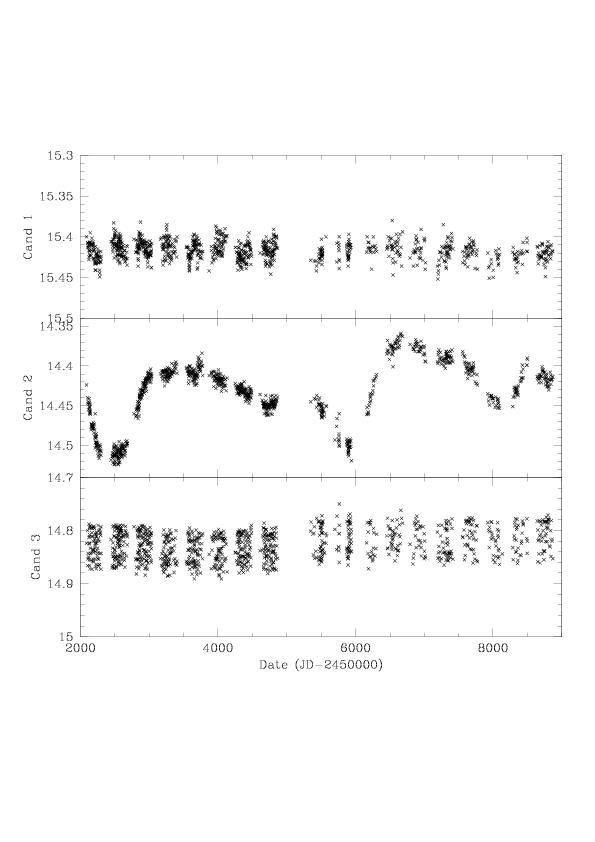}
    \caption{OGLE III and IV data for the three candidates for \sw{} discussed in this work. \textcolor{black} {The gap in the data around TJD 5000 indicates the transition from OGLE III to OGLE IV}. The vertical scale in each case is the I-band magnitude. Candidate 2 is here proposed to be the correct optical counterpart. }
    \label{fig:3panels}
\end{figure*}

The OGLE project \citep{Udalski2015}  provides long term I-band photometry with a cadence of 1-3 days. 

Fig \ref{fig:fc} is an OGLE I band image and it can be seen that there are 3 objects close to the position of the X-ray source. The northernmost one, labelled Candidate 1 in the figure, is [MA93] 302 \citep{ma93} which has previously been proposed as the counterpart to this X-ray source \citep{kennea2018,hs2016}. However, the new much-improved X-ray position from this work conclusively rules this out now. The southernmost object, Candidate 3, is probably also ruled out as it lies outside the 90\% uncertainty radius of the X-ray position.
Long term data from all three candidates are presented in Fig \ref{fig:3panels}. 

It is therefore concluded that, based upon position, that Candidate 2 is the most likely optical counterpart to \sw{}.

\subsubsection{Candidate 1}

For Candidate 1, [MA93] 302 \citep{ma93}, the OGLE identifications are SMC 103.5.17819 and SMC 720.20.4879D for OGLE III and IV respectively. The OGLE III \& IV data were investigated for any periodic behaviour which might indicate the binary period of the system. Lomb-Scargle techniques were applied \textcolor{black} {to the de-trended data} and periods investigated over the range : 1.1  -- 1000d. No significant periodic features were revealed. In fact over the 18 year period covered by the OGLE observations no fluctuations greater than 0.05 magnitudes in the I-band were observed. A behaviour pattern very different to that seen in almost all other Be stars in \bexrb ~systems.

\subsubsection{Candidate 2}

For Candidate 2 the OGLE identifications are SMC 103.5.17775 and SMC 720.20.4843D for OGLE III and IV respectively.  It is difficult to obtain the brightness of this object at different wavebands from the literature because of the danger of source confusion with Candidate 3. However, OGLE has observed this object also in V band during OGLE III and provides average values of V= 14.65 and I=14.45. \textcolor{black} {It is clear that the OGLE data for this source show a long-term modulation often characteristic of Be stars. This slow modulation is related to the growth and decay of the circumstellar disk which makes a significant contribution to the I-band light.}

\begin{figure}

	\includegraphics[width=8cm,angle=-0]{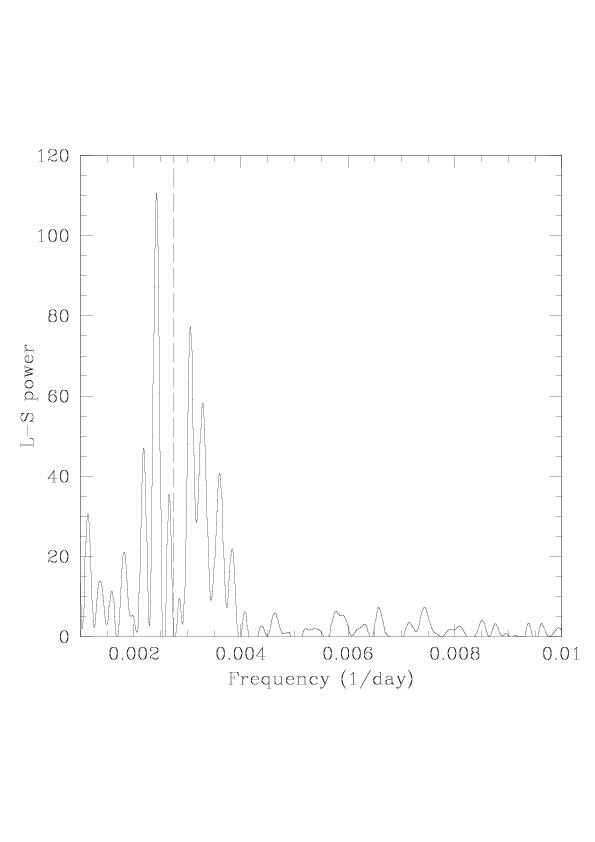}
    \caption{Lomb-Scargle timing analysis of the OGLE data for Candidate 2. The peak is at a period of $413\pm7$d; the dashed line shows the frequency corresponding to a period of 1 year.}
    \label{fig:ls}
\end{figure}

\textcolor{black} {The OGLE III \& IV data were de-trended wit a polynomial function and investigated for any periodic behaviour which might indicate the binary period of the system. Lomb-Scargle techniques were applied and periods investigated over the range : 1.1 -- 1000d. The former, OGLE III showed no significant peaks in the power spectrum, but the latter, OGLE IV, revealed a strong peak at 413$\pm$7d.} To check that this feature was not associated in any way with a window function such as the the annual sampling pattern, the simultaneously acquired OGLE data for Candidate 3 were also analysed in the same manner and then the two power spectra were subtracted from each other. The result is shown in Fig \ref{fig:ls}. The strong peak at 413d is unaffected by this process indicating that it was unrelated to the data structure. It can also be seen from this figure that the 413d peak is well separated from any structure associated with 1 year that may have survived the subtraction process.

To investigate the nature of the 413d modulation the OGLE data were then folded  with an ephemeris 

$T_0$ = 2086.90 + n(413.22) TJD. 

\begin{figure}

	\includegraphics[width=8cm,angle=-0]{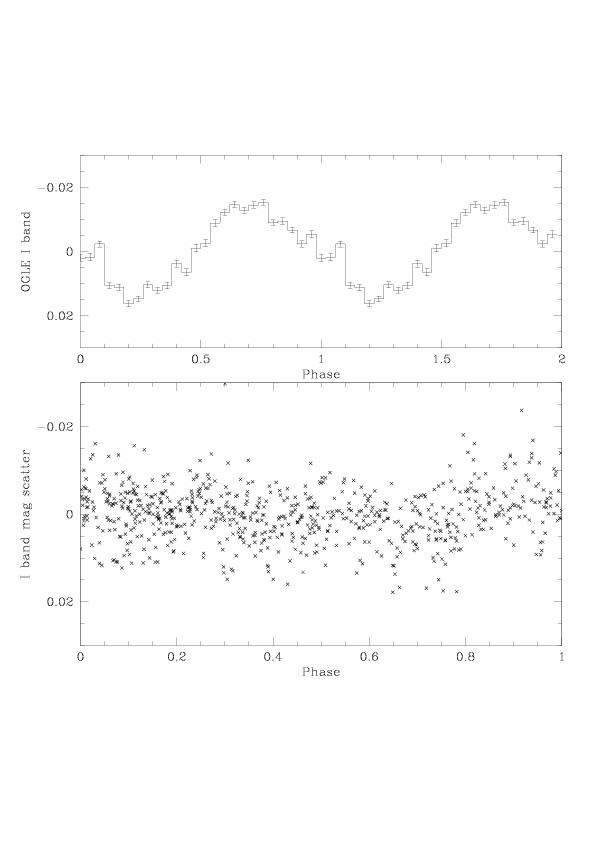}
    \caption{OGLE data for Candidate 2 folded at a period of 413d. \textcolor{black}{Top panel shows the binned data repeated over 2 cycles. Lower panel shows the scatter of the data points over just one cycle. In both cases the data have been de-trended with a polynomial fit and the fit then subtracted from the data.}}
    \label{fig:fold}
\end{figure}

The result is shown in Fig \ref{fig:fold} and reveals a sinusoidal type of modulation with an amplitude of $\sim$0.06 mags. This is further discussed below. 

\subsubsection{Candidate 3}

\begin{figure}
\hspace{0.1cm}
	\includegraphics[width=8.5cm,angle=-0]{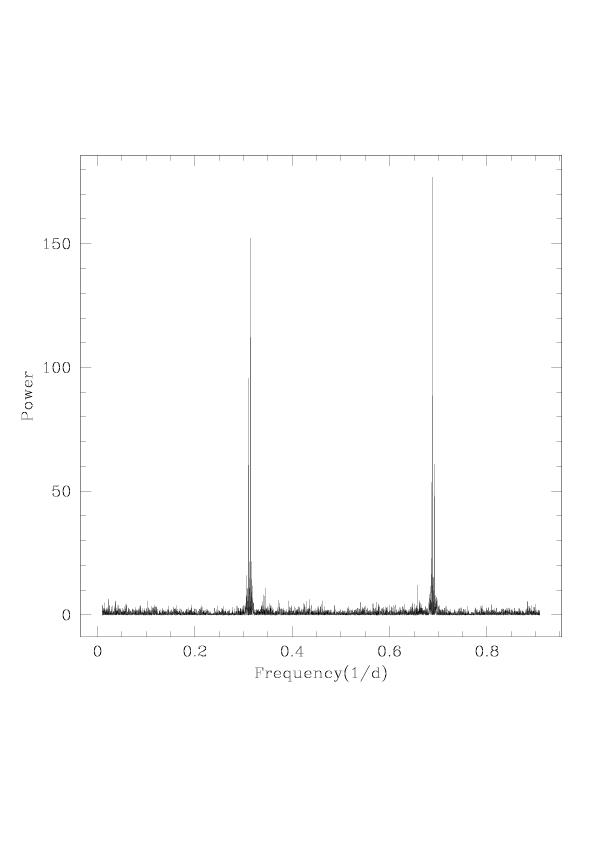}
    \caption{Power spectrum of the OGLE IV data for Candidate 3 over the period range 1--100d. }
    \label{fig:ls3}
\end{figure}

\begin{figure}
\hspace{0.1cm}
	\includegraphics[width=8.5cm,angle=-0]{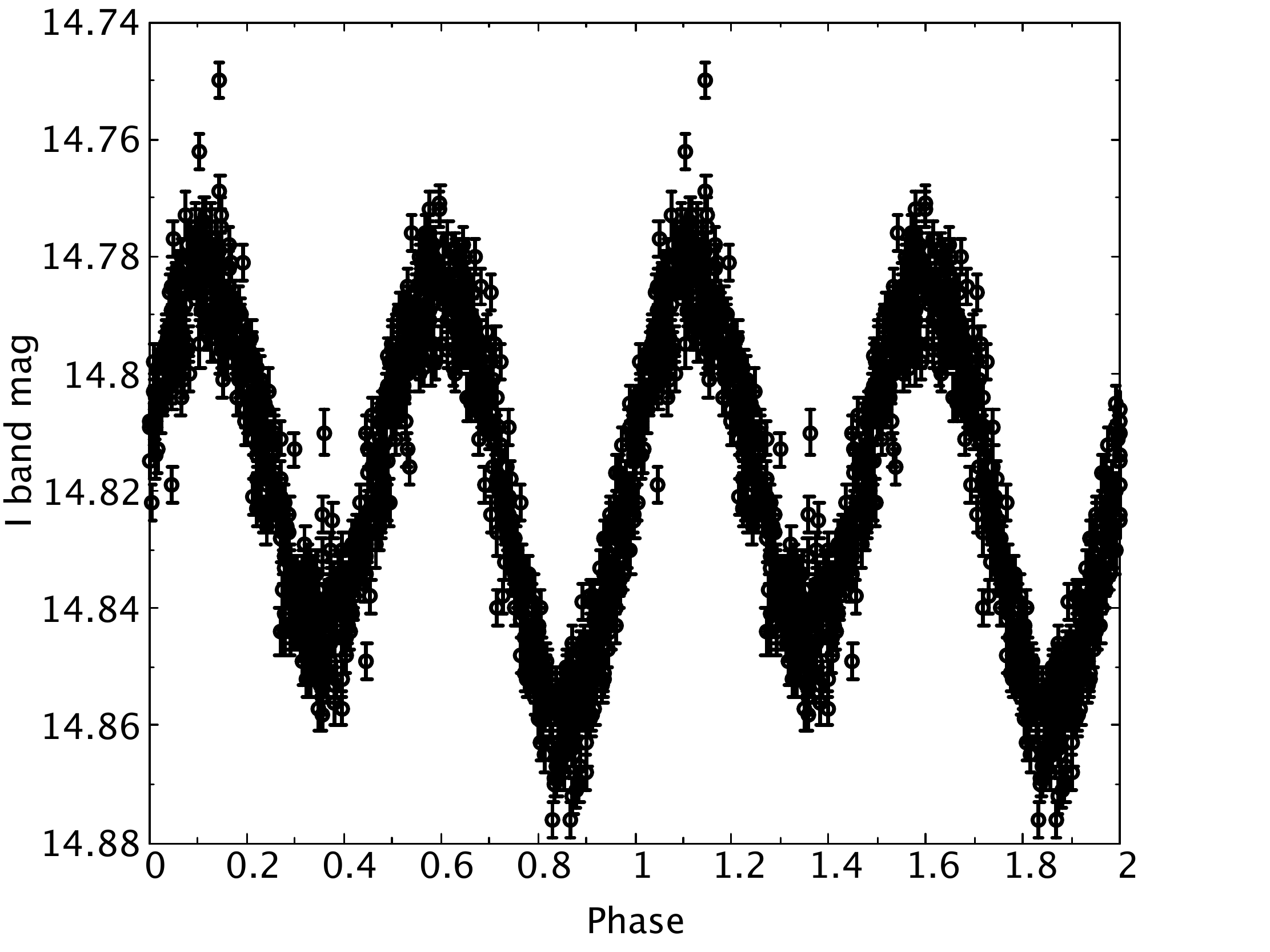}
    \caption{OGLE III and IV data for Candidate 3 folded at the period of 2.902d. }
    \label{fig:ogfold}
\end{figure}

\begin{figure}
\hspace{0.1cm}
	\includegraphics[width=8.5cm,angle=-0]{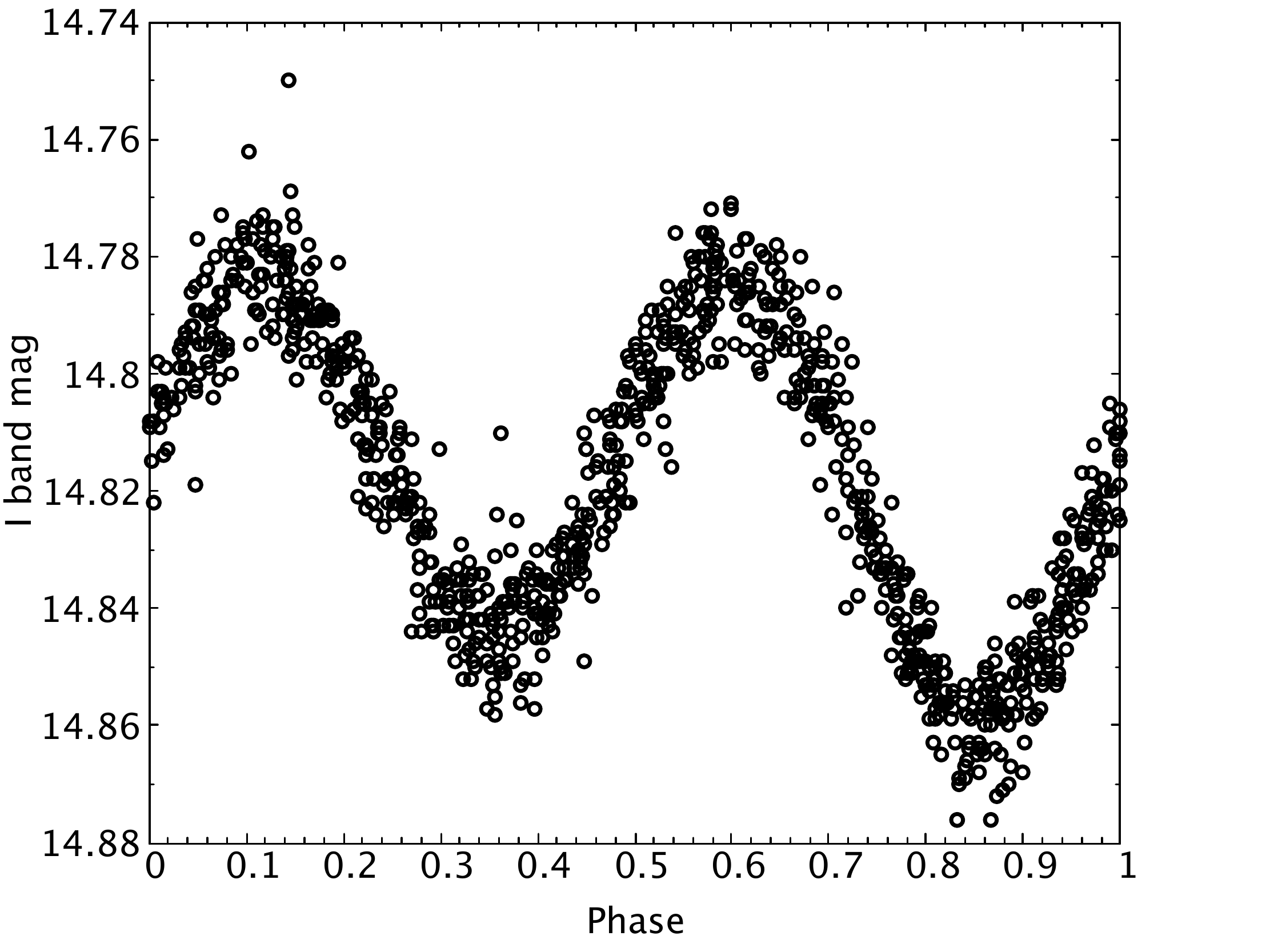}
    \caption{\textcolor{black}{OGLE III and IV data for Candidate 3 folded at the period of 2.902d without the error bars and only showing one cycle.} }
    \label{fig:ogfold2}
\end{figure}

For Candidate 3 the OGLE identifications are SMC 103.5.17780 and SMC 720.20.4842D for OGLE III and IV respectively. Lomb-Scargle analysis of these data reveals strong peaks at 1.4512d and 3.188d \textcolor{black}{ - see Fig \ref{fig:ls3}}. The latter period is interpreted as the beat period between the former period and the \textcolor{black} {sidereal daily period}. More careful inspection of the OGLE data reveal that the true period is, in fact, 2.902d with the folded data revealing an almost double sinusoidal profile - \textcolor{black} {see Figs \ref{fig:ogfold} and \ref{fig:ogfold2}}. 

\textcolor{black}{This system is very likely an ellipsoidal O-type star in a
binary system. The secondary is perhaps a compact object, but more likely a non-interacting companion (at least the OGLE data show no evidence for interacting behaviour), maybe a much fainter lower mass star. Several such systems are seen in the SMC \citep{pawlak2016}.}

\subsection{SALT}

\begin{figure}
\hspace{0.07cm}
	\includegraphics[width=0.95\columnwidth]{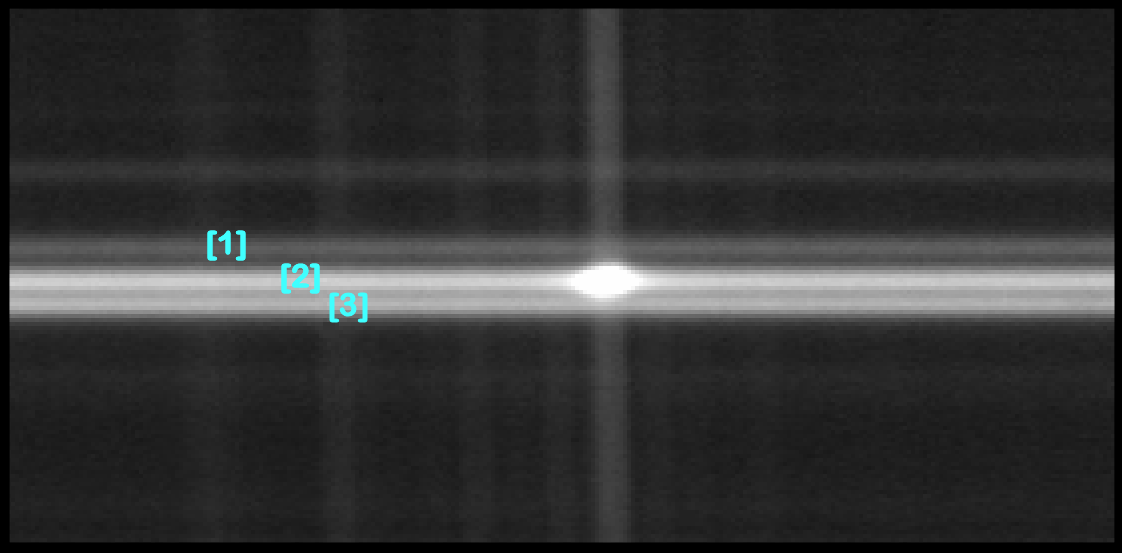}
    \caption{The region around H-alpha on the 2-D image taken with SALT. The spectral traces of candidates 1, 2 and 3 shown in Fig. \ref{fig:fc} can been seen in the centre of the image, running from top to bottom respectively. The H-alpha emission reported in section 2.3 can be clearly associated with candidate 2.}
    \label{fig:salt-trace}
\end{figure}

The three stars discussed above were observed with the Southern African Large Telescope (SALT) to help confirm Candidate 2 as the true optical counterpart to \sw{}. A single broadband spectrum was taken on 22 December 2020 as part of the SALT Transients large science programme. The observation was made with the Robert Stobie Spectrograph (RSS) in standard long-slit mode using the PG0900 grating and a 2" slit. The exposure time was 300\,s. The primary data reduction steps (which include overscan correlation, bias subtraction, gain and amplifier cross-talk corrections) were performed using the SALT science pipeline (\citealt{2010SPIE.7737E..25C}), while the secondary steps (wavelength calibration, background subtraction and extraction of the one-dimensional spectrum) were performed using standard IRAF routines. A second 300\,s exposure was made on the same night with the same setup, but unfortunately the seeing worsened causing the three target traces to become blurred, making this observation unusable. Fig. \ref{fig:salt-trace} shows part of the 2-D image obtained with SALT during the first observation. The spectral traces of Candidates 1, 2 and 3 can been seen in the centre of the image, running from top to bottom respectively. The bright emission is H-alpha, which can be clearly associated with Candidate 2. This is further strong evidence that Candidate 2 is the true counterpart - an emission line star.

\begin{figure*}
	\includegraphics[width=2.2\columnwidth]{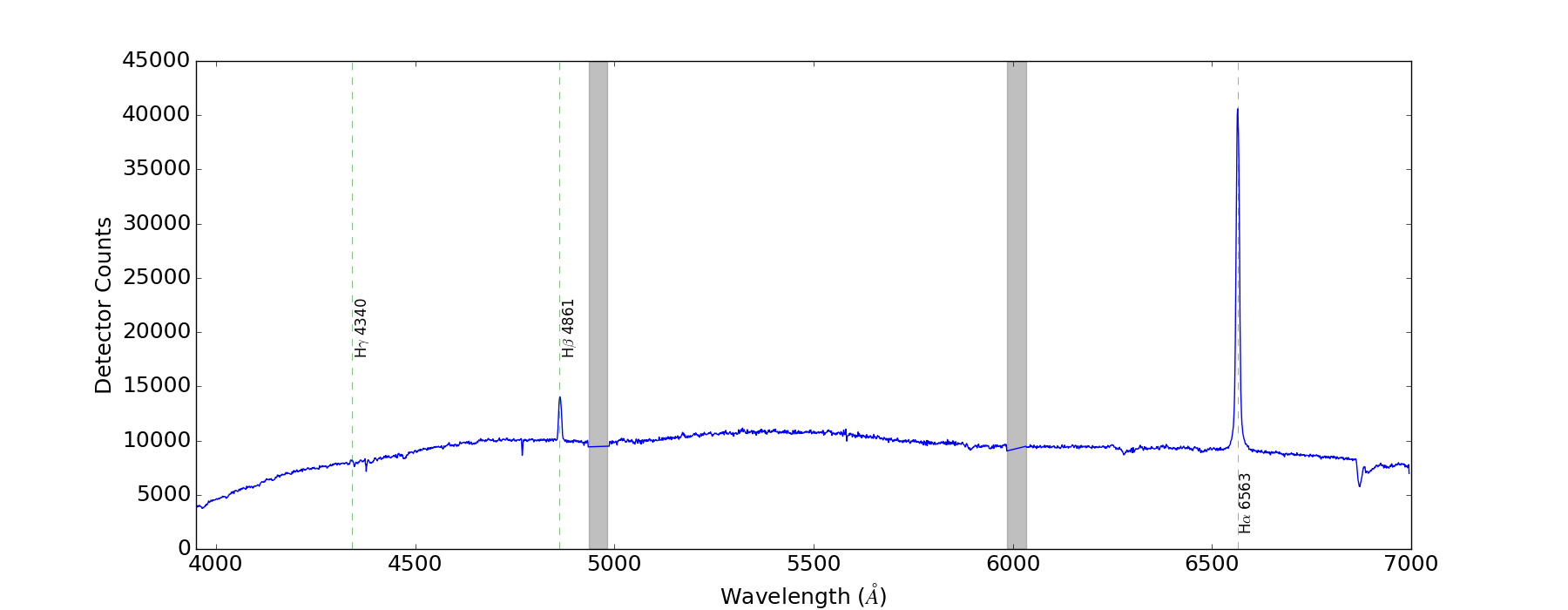}
	\includegraphics[width=2.2\columnwidth]{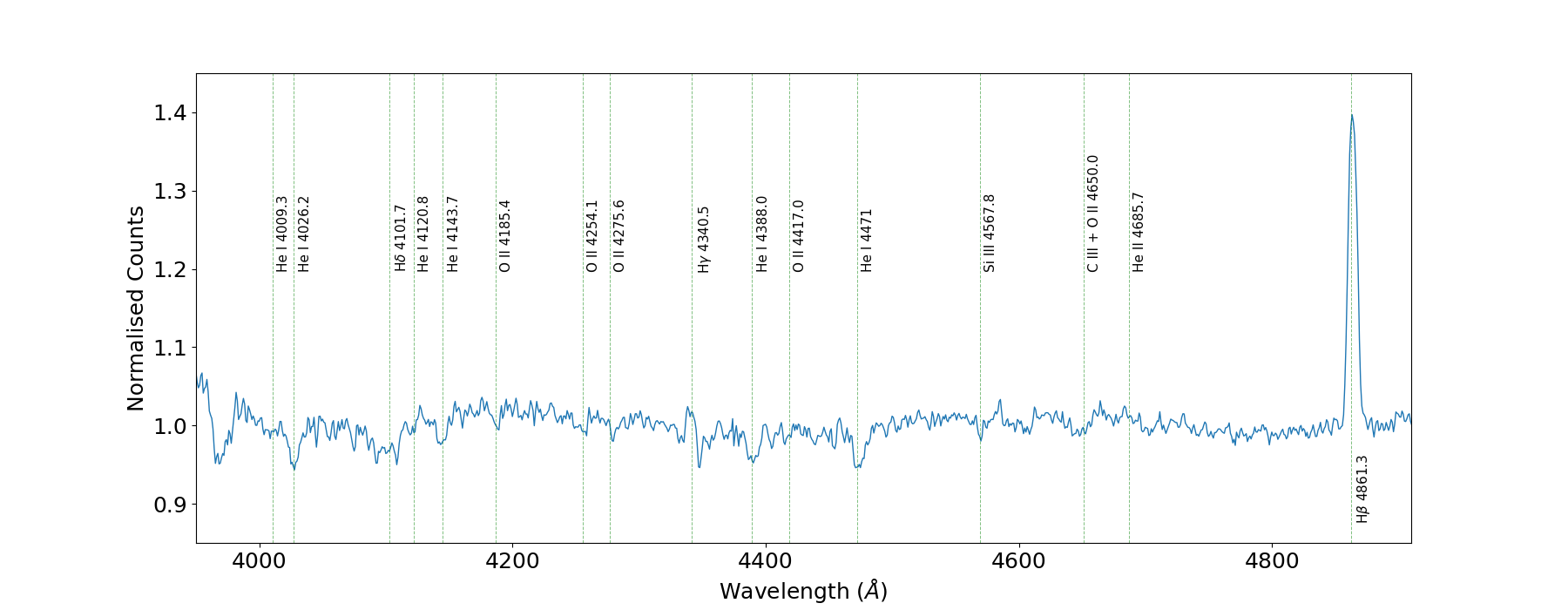}
    \caption{Top: SALT spectrum of candidate 2, with the Balmer lines labelled. Data taken on 20201222. The chip gaps around 5000~$\text{\AA}$ and 6000~$\text{\AA}$ have been interpolated over (highlighted in grey). Bottom: Normalised blue end section of the spectrum.}
    \label{fig:salt-full}
\end{figure*}

The close proximity of the three stars makes spectral extraction difficult. A simple projection across the 3 traces in Fig. \ref{fig:salt-trace} shows significant overlap between Candidates 2 and 3, and moderate overlap between Candidates 1 and 2. The spectra of all three stars were extracted as well as possible with these restrictions in mind. Due to its faintness, the spectrum of Candidate 1 showed significant contamination from Candidate 2. Strong H$\alpha$ emission is present in Candidate 1, which we believe to be contamination from Candidate 2. Aside from the contamination, Candidate 1 is very red. So much so that the trace is completely lost blue-ward of around 5000~$\text{\AA}$, making characterisation impossible. We are, however, highly confident that this star (previously identified in the literature as the optical counterpart to \sw) cannot be associated with the X-ray source.

The spectrum of Candidate 3 is much cleaner than Candidate 1, despite being closer to Candidate 2. This is likely due to its brightness being more comparable to Candidate 2. There is moderate H$\alpha$ emission present, which can, again, be explained as contamination from Candidate 2. The overall spectrum of Candidate 3 is very blue and shows strong Balmer and He I lines in absorption. We can also identify several weak lines of ionised oxygen and helium, suggesting that Candidate 3 is a late O-type star. Combined with the presence of a 2.9\,d period, this makes Candidate 3 an intriguing object. However, we suggest that it is unlikely to be the true counterpart of the X-ray source as its position lies significantly outside the \swift\ 90\% uncertainty.

Fig. \ref{fig:salt-full} shows the full SALT spectrum of Candidate 2 (top) and a zoom-in to the blue end (bottom). Immediately obvious is the presence of extremely strong H$\alpha$ emission (as can be seen in the 2-D image - see Fig. \ref{fig:salt-trace} ), as well as weaker emission in H$\beta$ and H$\gamma$, and a rather featureless continuum. This is indicative of an early-type star with a circumstellar disc, providing strong evidence that this system is definitely a \bexrb\ in the SMC. The Balmer line emission is discussed further in Section 3.2, alongside the spectral classification and limitations of our analysis based on the likely contamination of the spectrum from Candidate 3.

\begin{figure*}

	\includegraphics[width=16cm,angle=-0]{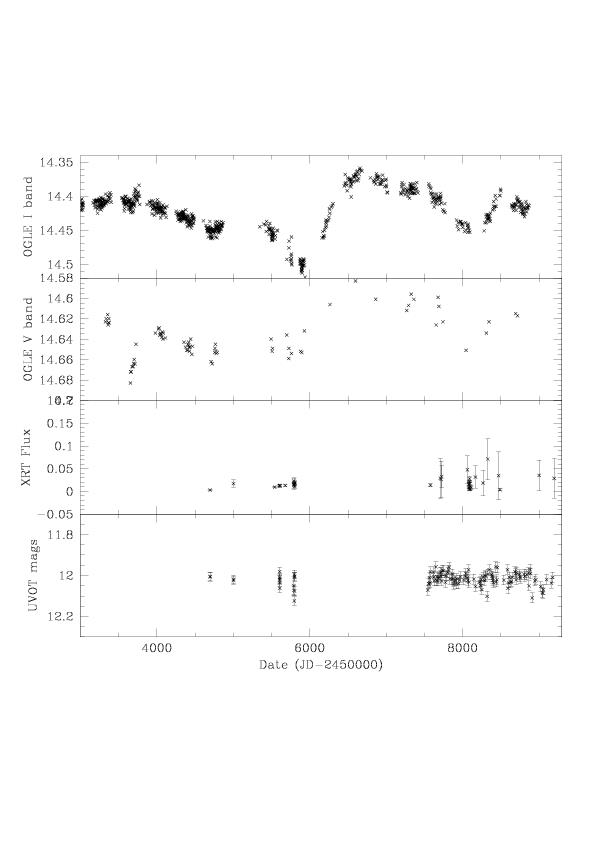}
    \caption{Candidate 2 observational data. Top panel : OGLE I-band data; second panel : OGLE V-band data; third panel : \swift\ X-ray 2-10 keV; lowest panel : \swift\ UVOT data in the uvw1 filter. XRT and UVOT observations before TJD 7500 are from historical \swift\ observations prior to the start of the S-CUBED project.}
    \label{fig:xouv}
\end{figure*}

\section{Discussion}

Since it is extremely likely that Candidate 2 is the correct optical counterpart to \sw{} the discussion is restricted to just that object.

\subsection{Long term variability}

Fig \ref{fig:xouv} shows the combined data set for this source during the period JD 2453000-2459000: two OGLE bands (I \& V), \swift\ X-ray 2-10 keV, and \swift\ UV (filter uvw1 centred on 2600\AA). The behaviour seen in the I band is characteristic of most Be stars found in \bexrb\ systems, showing long term fluctuations arising from the alterations in the extent of the circumstellar disks around the OB star. The V band generally follows the I band behaviour, but at a lower amplitude ($\sim$0.07 mags compared to $\sim$0.15 mags in the I band). This is understandable as at least 50\% of the V band flux is arising directly from the OB star and is much more stable. This is further supported by the negligible fluctuations in the lowest panel - that of the UV band. The X-ray flux shows little variation and is probably always close to the detection limit of S-CUBE. Such modest X-ray fluxes are characteristic of \bexrb\ with long, low-eccentricity orbital periods.

\begin{figure*}%
    \centering
    \subfloat[ ]{{\includegraphics[width=8cm]{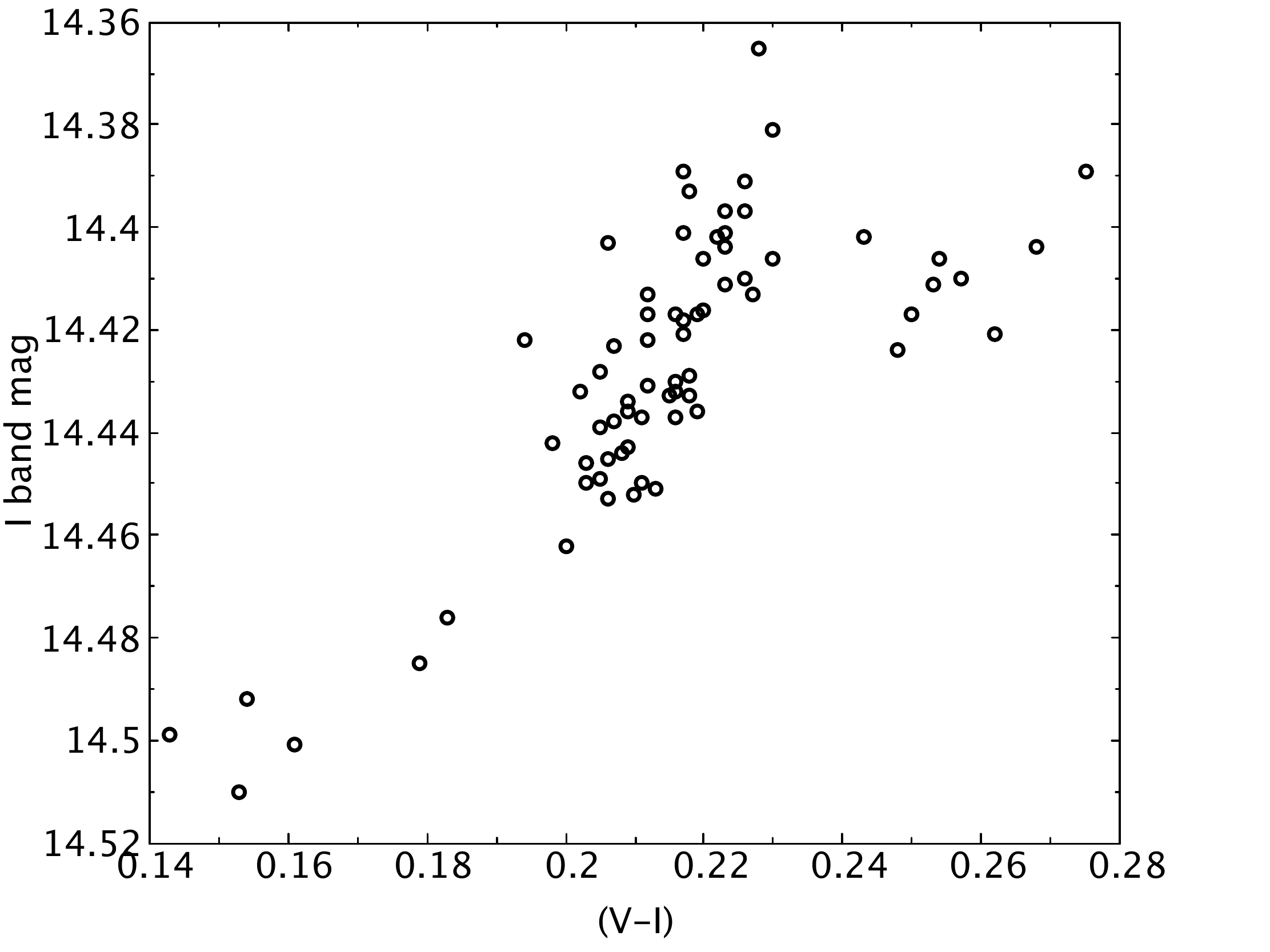} }}%
    \qquad
    \subfloat[ ]{{\includegraphics[width=8cm]{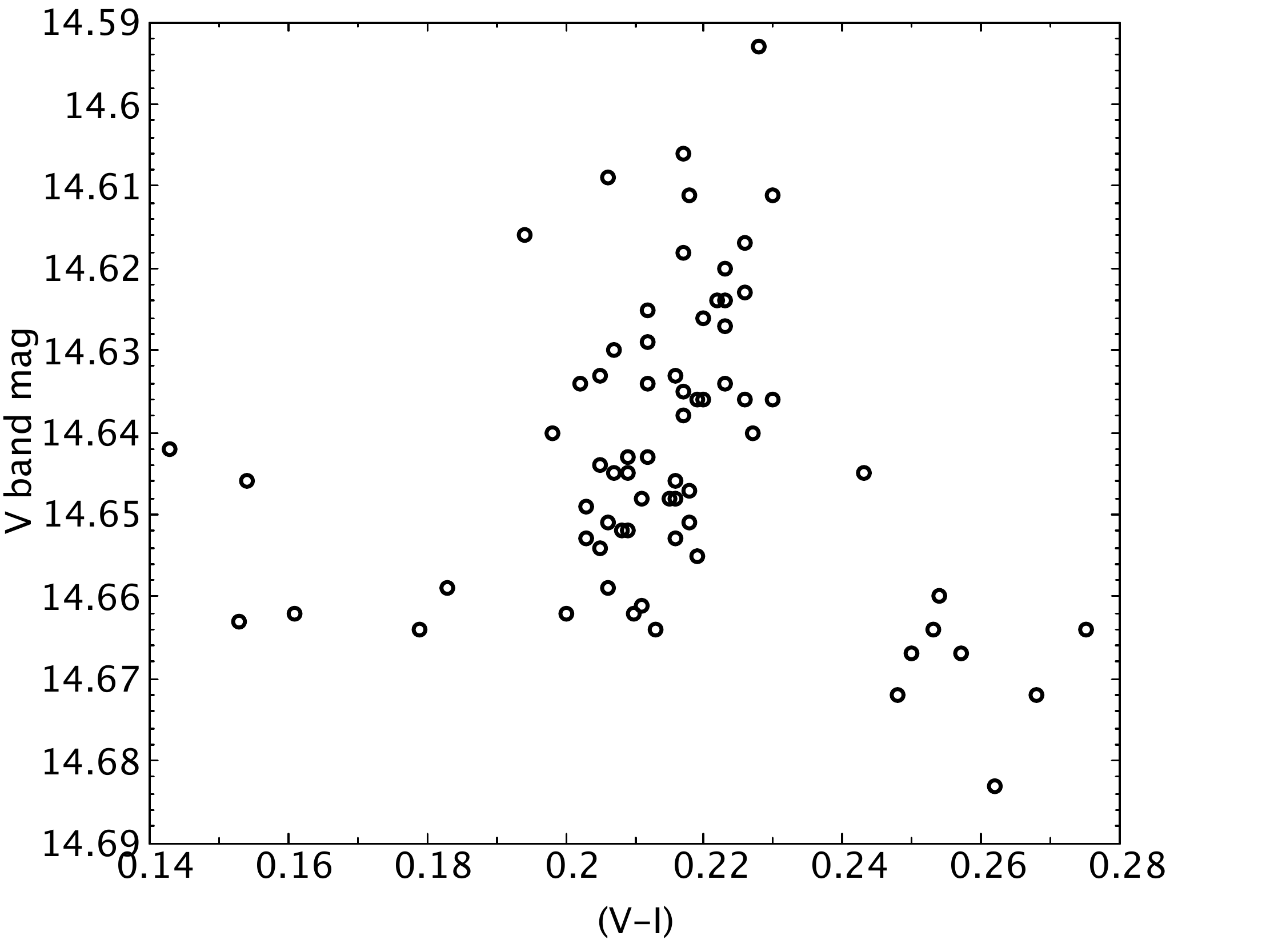} }}%
    \caption{$(V-I)-I$ (left) and $(V-I)-V$ (right) colour-magnitude plots from OGLE data. }%
    \label{fig:col_mag}%
\end{figure*}

Fig \ref{fig:col_mag} shows the colour variation seen in the disk over the time of the OGLE data JD 2453326 - 2458745 when simultaneous V and I band observations were performed. During this period the photometric flux was showing a steady decline accompanied by a dramatic dip during the period 3600 < TJD < 3800, which is more pronounced in the V-band. This is possibly an indicator of loss of the inner regions of the disc, from where the bluer emission originates. At $\sim$TJD 5500, the system underwent a significant drop in I-band flux while the V-band magnitude remained constant, suggesting disc-loss in the outer regions during this time. 

Fig \ref{fig:col_mag} (a) reveals general trend of the colour, being redder when the system is brighter. This possibly indicates a low/intermediate inclination angle of the disc with a flaring circumstellar disk structure partially obscuring the bluer light directly from the star. The prominent dip in the V-band emission during 3600 < TJD < 3800 results in colour variations happening in two discontinuous ranges, 0.14$\le$(V-I)$\le$0.23 and 0.24$\le$(V-I)$\le$0.28. In the second colour range (0.24$\le$(V-I)$\le$0.28), the general trend of the colour being redder when brighter persists in the (V-I)-I plot (Fig \ref{fig:col_mag} (a)), while it is less apparent in the (V-I)-V plot (Fig \ref{fig:col_mag} (b)). The (V-I)-V distribution also shows more scatter, suggesting more significant changes occurring in the inner disc regions

\subsection{Optical spectrum}

The equivalent widths of the H$\alpha$ and H$\beta$ lines measured from the spectrum of Candidate 2 are -40.6 $\pm$ 0.4~$\text{\AA}$ and -3.6 $\pm$ 0.2~$\text{\AA}$, respectively. The H$\alpha$ equivalent width is significantly larger than the average seen in the \bexrb\ population. Using the proposed relationship between H$\alpha$ equivalent width and orbital period \citep{rfc1997, ck2015}, the value measured suggests an orbital period of the order 200-400 days. This is consistent with the 413d period seen in the OGLE data.

Besides the strong Balmer emission, one can see several lines of neutral helium in absorption in the spectrum (Fig. \ref{fig:salt-full}). There is no evidence for any ionised helium, meaning the spectral type must be later than B\,0.5. It is difficult to determine which metal lines are present, and the strength of these lines, due to the low signal-to-noise in the blue part of the spectrum. Mg II seems to be absent, or at least very weak, indicating the spectral type may be earlier than B3. There is good evidence for OII lines, which seem stronger than either of the Si IV lines that may be present around 4100~$\text{\AA}$. This would indicate a spectral type of B\,1.5 or earlier. However, as mentioned earlier, there is potential contamination from Candidate 3 to consider. The strong He I and ionised oxygen lines seen in Candidate 3 may be contributing signal to our target spectrum, making a precise classification difficult. Because of this, we are not confident enough to be able to distinguish between a B2 and B3 late limit for the spectral type. Thus, we suggest a spectral type between B\,1--3, though we stress that a higher quality spectrum taken in better seeing conditions is needed to confirm this, and to provide a finer classification.

Given that the metal lines in the spectrum are often not obvious, we are limited to using the HeI\,4121/HeI\,4143 ratio as one approach to determine the luminosity class. This line ratio increases towards more luminous stars, suggesting that our object has a luminosity class of III-V. This result may be compared to that obtained by using the absolute magnitude of the star in the V-band with the distance modulus for the SMC \citep{graczyk2014}. Using the tables of \citet{weg2006}, a V-band magnitude of 14.6 favours a luminosity class of III-IV for a spectral type of B1-3. A star of luminosity class V would need to have an earlier spectral type of $\sim$O9 for this magnitude.

Based on the information provided by the optical spectrum and magnitude of the star, and the limitations of the classification discussed above, our best estimate of the spectral type of this system is B\,1--3 III-IVe.

Finally we note that not only is Candidate 1 ([MA93] 302) positionally inconsistent as the optical counterpart to \sw{} but it is not even an early type star. It is quite possible that light from nearby Candidate 2 led to its incorrect cataloguing as an early type emission line star by \cite{ma93}.

\subsection{Optical periodic modulation}

The optical modulation at a period of 413d shown in Fig \ref{fig:fold} is interpreted as the binary period of the system and is very sinusoidal in shape. This is very different from the sharp outburst profiles seen in other long period \bexrb\ systems such as SXP 756 with its similar orbital period of 394d \citep{ce2004}. The most obvious interpretation of these differences must lie in the eccentricity of the neutron star orbit and the manner in which it interacts with the circumstellar disk of the Be star. These disks are assumed to be optically thick, so any variation seen in the OGLE I-band data must be interpreted as a change in the surface area of the disk. Regular changes being brought about by the approach of the neutron star at periastron passage. Hence the more gentle long term changes observed here in \sw{} must surely indicate a very low eccentricity orbit for the neutron star. This is further supported by the small depth of the observed modulation: ~$\sim$0.06 mags in \sw{} compared to $\sim$0.5 mags in SXP 756. 

If this 413d period is interpreted as the binary period, then an estimate of the likely spin period of the neutron star may be inferred from the Corbet Diagram \citep{corbet1984}. This is a very long binary period for \bexrb\ systems and places \sw\ right at the edge of the known distribution, it also suggests an exceptionally long pulse period in the region of $\ge$1000s. Such a long period is hard to measure since an extended X-ray observation of many ks would be required whilst the source was reasonably bright. An example of a source that lies in this extreme zone on the Corbet Diagram is SAX J2239.3+6116 \citep{zand2001} in the Galaxy with an orbital period of 262d and a pulse period of 1247s. Within the SMC, there are three systems with proposed pulse periods in excess of 1ks : SXP 1062, SXP 1323 and SXP 4693 (= CXOU J005446.3-722523). The latter is most exceptional and has no known binary period \citep{laycock2010}. Such long pulse period systems tend not to be bright X-ray sources, making the detection of pulse \& binary periods challenging - see \cite{laycock2010} for a discussion on such long pulse period systems. Thus it is valuable to add \sw{} to the list of these exceptional \bexrb\ systems.

\section{Conclusions}

Reported here is the confirmation that \sw{} is a \bexrb\ system in the SMC. As a result of many observations arising from the S-CUBED project a precise position has been determined for this X-ray source. This has enabled the correct identification of the optical counterpart, resolving a long-standing confusion in a crowded stellar field. The long-term OGLE data has made it possible to detect the probable binary period for this system (413d) making it one of the longest known for a \bexrb\ system. Such long period systems are valuable in testing the limits of the correlation between spin \& orbital periods seen in the Corbet diagram, with possible implications for accretion modes in these well separated binary systems. 

\section{Data availability}

All X-ray data are freely available from the Swift archive. The other data underlying this article will be shared on any reasonable request to the corresponding author.

\section*{Acknowledgements}

The OGLE project has received funding from the National Science Centre, Poland, grant MAESTRO 2014/14/A/ST9/00121 to AU. PAE acknowledges UKSA support. IMM acknowledges support from the National Research Foundation (NRF) and the University of Cape Town (UCT).
This work made use of data supplied by the UK Swift Science Data Centre at the University of Leicester.
LJT acknowledges support from the National Research Foundation (NRF), South Africa and the SALT consortium. Some of the observations reported in this paper were obtained with the Southern African Large Telescope, as part of the Large Science Programme on transients 2018-2-LSP-001 (PI: Buckley).

\textcolor{black}{We are grateful to the anonymous referee for several helpful suggestion to improve the paper.}




\bibliographystyle{mnras}
\bibliography{references}

\begin{thebibliography}{}
\makeatletter
\relax
\def\mn@urlcharsother{\let\do\@makeother \do\$\do\&\do\#\do\^\do\_\do\%\do\~}
\def\mn@doi{\begingroup\mn@urlcharsother \@ifnextchar [ {\mn@doi@}
  {\mn@doi@[]}}
\def\mn@doi@[#1]#2{\def\@tempa{#1}\ifx\@tempa\@empty \href
  {http://dx.doi.org/#2} {doi:#2}\else \href {http://dx.doi.org/#2} {#1}\fi
  \endgroup}
\def\mn@eprint#1#2{\mn@eprint@#1:#2::\@nil}
\def\mn@eprint@arXiv#1{\href {http://arxiv.org/abs/#1} {{\tt arXiv:#1}}}
\def\mn@eprint@dblp#1{\href {http://dblp.uni-trier.de/rec/bibtex/#1.xml}
  {dblp:#1}}
\def\mn@eprint@#1:#2:#3:#4\@nil{\def\@tempa {#1}\def\@tempb {#2}\def\@tempc
  {#3}\ifx \@tempc \@empty \let \@tempc \@tempb \let \@tempb \@tempa \fi \ifx
  \@tempb \@empty \def\@tempb {arXiv}\fi \@ifundefined
  {mn@eprint@\@tempb}{\@tempb:\@tempc}{\expandafter \expandafter \csname
  mn@eprint@\@tempb\endcsname \expandafter{\@tempc}}}

\bibitem[\protect\citeauthoryear{{Burrows} et~al.,}{{Burrows}
  et~al.}{2005}]{burrows05}
{Burrows} D.~N.,  et~al., 2005, \mn@doi [\ssr] {10.1007/s11214-005-5097-2},
  \href {http://adsabs.harvard.edu/abs/2005SSRv..120..165B} {120, 165}

\bibitem[\protect\citeauthoryear{{Coe} \& {Edge}}{{Coe} \&
  {Edge}}{2004}]{ce2004}
{Coe} M.~J.,  {Edge} W.~R.~T.,  2004, \mn@doi [\mnras]
  {10.1111/j.1365-2966.2004.07696.x}, \href
  {https://ui.adsabs.harvard.edu/abs/2004MNRAS.350..756C} {350, 756}

\bibitem[\protect\citeauthoryear{{Coe} \& {Kirk}}{{Coe} \&
  {Kirk}}{2015}]{ck2015}
{Coe} M.~J.,  {Kirk} J.,  2015, \mn@doi [\mnras] {10.1093/mnras/stv1283}, \href
  {https://ui.adsabs.harvard.edu/abs/2015MNRAS.452..969C} {452, 969}

\bibitem[\protect\citeauthoryear{{Corbet}}{{Corbet}}{1984}]{corbet1984}
{Corbet} R.~H.~D.,  1984, \aap, \href
  {https://ui.adsabs.harvard.edu/abs/1984A&A...141...91C} {141, 91}

\bibitem[\protect\citeauthoryear{{Crawford} et~al.,}{{Crawford}
  et~al.}{2010}]{2010SPIE.7737E..25C}
{Crawford} S.~M.,  et~al., 2010, in {Silva} D.~R.,  {Peck} A.~B.,   {Soifer}
  B.~T.,  eds,  Society of Photo-Optical Instrumentation Engineers (SPIE)
  Conference Series Vol. 7737, Observatory Operations: Strategies, Processes,
  and Systems III. p. 773725, \mn@doi{10.1117/12.857000}

\bibitem[\protect\citeauthoryear{{Evans} et~al.,}{{Evans}
  et~al.}{2009}]{Evans09}
{Evans} P.~A.,  et~al., 2009, \mn@doi [\mnras]
  {10.1111/j.1365-2966.2009.14913.x}, \href
  {http://adsabs.harvard.edu/abs/2009MNRAS.397.1177E} {397, 1177}

\bibitem[\protect\citeauthoryear{{Evans} et~al.,}{{Evans}
  et~al.}{2014}]{Evans14}
{Evans} P.~A.,  et~al., 2014, \mn@doi [\apjs] {10.1088/0067-0049/210/1/8},
  \href {http://adsabs.harvard.edu/abs/2014ApJS..210....8E} {210, 8}

\bibitem[\protect\citeauthoryear{{Gehrels} et~al.,}{{Gehrels}
  et~al.}{2004}]{gehrels04}
{Gehrels} N.,  et~al., 2004, \mn@doi [\apj] {10.1086/422091}, \href
  {https://ui.adsabs.harvard.edu/abs/2004ApJ...611.1005G} {611, 1005}

\bibitem[\protect\citeauthoryear{{Graczyk} et~al.,}{{Graczyk}
  et~al.}{2014}]{graczyk2014}
{Graczyk} D.,  et~al., 2014, \mn@doi [\apj] {10.1088/0004-637X/780/1/59}, \href
  {https://ui.adsabs.harvard.edu/abs/2014ApJ...780...59G} {780, 59}

\bibitem[\protect\citeauthoryear{{Haberl} \& {Sasaki}}{{Haberl} \&
  {Sasaki}}{2000}]{hs2000}
{Haberl} F.,  {Sasaki} M.,  2000, \aap, \href
  {https://ui.adsabs.harvard.edu/abs/2000A&A...359..573H} {359, 573}

\bibitem[\protect\citeauthoryear{{Haberl} \& {Sturm}}{{Haberl} \&
  {Sturm}}{2016}]{hs2016}
{Haberl} F.,  {Sturm} R.,  2016, \mn@doi [\aap] {10.1051/0004-6361/201527326},
  \href {https://ui.adsabs.harvard.edu/abs/2016A&A...586A..81H} {586, A81}

\bibitem[\protect\citeauthoryear{{Haberl}, {Eger}  \& {Pietsch}}{{Haberl}
  et~al.}{2008}]{hep2008}
{Haberl} F.,  {Eger} P.,   {Pietsch} W.,  2008, \mn@doi [\aap]
  {10.1051/0004-6361:200810100}, \href
  {https://ui.adsabs.harvard.edu/abs/2008A&A...489..327H} {489, 327}

\bibitem[\protect\citeauthoryear{{Kennea}, {Coe}, {Evans}, {Waters}  \&
  {Jasko}}{{Kennea} et~al.}{2018}]{kennea2018}
{Kennea} J.~A.,  {Coe} M.~J.,  {Evans} P.~A.,  {Waters} J.,   {Jasko} R.~E.,
  2018, \mn@doi [\apj] {10.3847/1538-4357/aae839}, \href
  {https://ui.adsabs.harvard.edu/abs/2018ApJ...868...47K} {868, 47}

\bibitem[\protect\citeauthoryear{{Laycock}, {Zezas}, {Hong}, {Drake}  \&
  {Antoniou}}{{Laycock} et~al.}{2010}]{laycock2010}
{Laycock} S.,  {Zezas} A.,  {Hong} J.,  {Drake} J.~J.,   {Antoniou} V.,  2010,
  \mn@doi [\apj] {10.1088/0004-637X/716/2/1217}, \href
  {https://ui.adsabs.harvard.edu/abs/2010ApJ...716.1217L} {716, 1217}

\bibitem[\protect\citeauthoryear{{Meyssonnier} \& {Azzopardi}}{{Meyssonnier} \&
  {Azzopardi}}{1993}]{ma93}
{Meyssonnier} N.,  {Azzopardi} M.,  1993, \aaps, \href
  {https://ui.adsabs.harvard.edu/abs/1993A&AS..102..451M} {102, 451}

\bibitem[\protect\citeauthoryear{{Orio}, {Zezas}, {Munari}, {Siviero}  \&
  {Tepedelenlioglu}}{{Orio} et~al.}{2007}]{orio2007}
{Orio} M.,  {Zezas} A.,  {Munari} U.,  {Siviero} A.,   {Tepedelenlioglu} E.,
  2007, \mn@doi [\apj] {10.1086/514806}, \href
  {https://ui.adsabs.harvard.edu/abs/2007ApJ...661.1105O} {661, 1105}

\bibitem[\protect\citeauthoryear{{Pawlak} et~al.,}{{Pawlak}
  et~al.}{2016}]{pawlak2016}
{Pawlak} M.,  et~al., 2016, \actaa, \href
  {https://ui.adsabs.harvard.edu/abs/2016AcA....66..421P} {66, 421}

\bibitem[\protect\citeauthoryear{{Reig}, {Fabregat}  \& {Coe}}{{Reig}
  et~al.}{1997}]{rfc1997}
{Reig} P.,  {Fabregat} J.,   {Coe} M.~J.,  1997, \aap, \href
  {https://ui.adsabs.harvard.edu/abs/1997A&A...322..193R} {322, 193}

\bibitem[\protect\citeauthoryear{{Roming} et~al.,}{{Roming}
  et~al.}{2005}]{Roming05}
{Roming} P.~W.~A.,  et~al., 2005, \mn@doi [\ssr] {10.1007/s11214-005-5095-4},
  \href {http://adsabs.harvard.edu/abs/2005SSRv..120...95R} {120, 95}

\bibitem[\protect\citeauthoryear{{Sasaki}, {Haberl}  \& {Pietsch}}{{Sasaki}
  et~al.}{2000}]{sasaki2000}
{Sasaki} M.,  {Haberl} F.,   {Pietsch} W.,  2000, \mn@doi [\aaps]
  {10.1051/aas:2000290}, \href
  {https://ui.adsabs.harvard.edu/abs/2000A&AS..147...75S} {147, 75}

\bibitem[\protect\citeauthoryear{{Scowcroft}, {Freedman}, {Madore}, {Monson},
  {Persson}, {Rich}, {Seibert}  \& {Rigby}}{{Scowcroft}
  et~al.}{2016}]{scowcroft2016}
{Scowcroft} V.,  {Freedman} W.~L.,  {Madore} B.~F.,  {Monson} A.,  {Persson}
  S.~E.,  {Rich} J.,  {Seibert} M.,   {Rigby} J.~R.,  2016, \mn@doi [\apj]
  {10.3847/0004-637X/816/2/49}, \href
  {https://ui.adsabs.harvard.edu/abs/2016ApJ...816...49S} {816, 49}

\bibitem[\protect\citeauthoryear{{Shtykovskiy} \& {Gilfanov}}{{Shtykovskiy} \&
  {Gilfanov}}{2005}]{sg2005}
{Shtykovskiy} P.,  {Gilfanov} M.,  2005, \mn@doi [\mnras]
  {10.1111/j.1365-2966.2005.09320.x}, \href
  {https://ui.adsabs.harvard.edu/abs/2005MNRAS.362..879S} {362, 879}

\bibitem[\protect\citeauthoryear{{Udalski}, {Szyma{\'n}ski}  \&
  {Szyma{\'n}ski}}{{Udalski} et~al.}{2015}]{Udalski2015}
{Udalski} A.,  {Szyma{\'n}ski} M.~K.,   {Szyma{\'n}ski} G.,  2015, \actaa,
  \href {https://ui.adsabs.harvard.edu/abs/2015AcA....65....1U} {65, 1}

\bibitem[\protect\citeauthoryear{{Wegner}}{{Wegner}}{2006}]{weg2006}
{Wegner} W.,  2006, \mn@doi [\mnras] {10.1111/j.1365-2966.2006.10549.x}, \href
  {https://ui.adsabs.harvard.edu/abs/2006MNRAS.371..185W} {371, 185}

\bibitem[\protect\citeauthoryear{{Willingale}, {Starling}, {Beardmore},
  {Tanvir}  \& {O'Brien}}{{Willingale} et~al.}{2013}]{Willingale2013}
{Willingale} R.,  {Starling} R.~L.~C.,  {Beardmore} A.~P.,  {Tanvir} N.~R.,
  {O'Brien} P.~T.,  2013, \mn@doi [\mnras] {10.1093/mnras/stt175}, \href
  {https://ui.adsabs.harvard.edu/abs/2013MNRAS.431..394W} {431, 394}

\bibitem[\protect\citeauthoryear{{in't Zand}, {Swank}, {Corbet}  \&
  {Markwardt}}{{in't Zand} et~al.}{2001}]{zand2001}
{in't Zand} J.~J.~M.,  {Swank} J.,  {Corbet} R.~H.~D.,   {Markwardt} C.~B.,
  2001, \mn@doi [\aap] {10.1051/0004-6361:20011512}, \href
  {https://ui.adsabs.harvard.edu/abs/2001A&A...380L..26I} {380, L26}

\makeatother
\end{thebibliography}

\label{lastpage}
\bsp	

\end{document}
